\documentclass[twocolumn,showpacs,aps,prl]{revtex4}
\usepackage{graphicx}
\usepackage{dcolumn}
\usepackage{amsmath}
\usepackage{latexsym}

\begin {document}

%{\underline {\bf Not for distribution please}}
%\vskip 0.2 cm

\title {Scaling forms for Relaxation Times of the Fiber Bundle model}
\author
{Chandreyee Roy, Sumanta Kundu and S. S. Manna}
\affiliation
{
\begin {tabular}{c}
Satyendra Nath Bose National Centre for Basic Sciences,
Block-JD, Sector-III, Salt Lake, Kolkata-700098, India
\end{tabular}
}
\begin{abstract}
      Using extensive numerical analysis of the Fiber Bundle Model with Equal Load Sharing dynamics we studied the 
   finite-size scaling forms of the relaxation times against the deviations of applied load per fiber from the 
   critical point. Our most crucial result is we have not found any $\ln (N)$ dependence of the average relaxation 
   time $\langle T(\sigma,N) \rangle$ in the precritical state. The other results are: (i) The critical load 
   $\sigma_c(N)$ for the bundle of size $N$ approaches its asymptotic 
   value $\sigma_c(\infty)$ as $\sigma_c(N) = \sigma_c(\infty) + AN^{-1/\nu}$. (ii) Right at the critical point the average 
   relaxation time $\langle T(\sigma_c(N),N) \rangle$ scales with the bundle size $N$ as: 
   $\langle T(\sigma_c(N),N) \rangle \sim N^{\eta}$ and this behavior remains valid within a small window of size 
   $|\Delta \sigma| \sim N^{-\zeta}$ around the critical point. (iii) When $1/N < |\Delta \sigma| < 100N^{-\zeta}$ the finite-size scaling 
   takes the form: $\langle T(\sigma,N) \rangle / N^{\eta} \sim {\cal G}[\{\sigma_c(N)-\sigma\}N^{\zeta}]$
   so that in the limit of $N \to \infty$ one has $\langle T(\sigma) \rangle \sim (\sigma - \sigma_c)^{-\tau}$.
   The high precision of our numerical estimates led us to verify that $\nu = 3/2$, conjecture that $\eta = 1/3$, $\zeta = 2/3$ and
   therefore $\tau = 1/2$.
\end{abstract}

\maketitle

      Fiber bundle models are used in Material Science to study the breakdown properties of materials in the form of a bundle
   composed of a large number of parallel massless elastic fibers \cite {Herrmann, Chakrabarti,Sornette,Sahimi,Bhattacharya}. 
   It is well known that the failure of the entire fiber bundle occurs at a critical value $\sigma_c$ of the applied load per 
   fiber and at this point the system undergoes a change from a state of local failure to a state of global failure. Consequently 
   $\sigma_c$ acts similar to the critical point of a phase transition and the behavior of the bundle around this point is associated 
   with all characteristics of critical phenomena. In this article we studied the relaxation behavior of fiber bundles at and
   very close to the critical point using extensive numerical simulations. We showed that away from the critical point the 
   relaxation times obey the usual finite size scaling theory. More interestingly we found that the amplitude of variation
   has no logarithmic dependence in the precritical regime as predicted in the mean field theory of fiber bundles 
   \cite {Pradhan1, Pradhan2}.

      The fiber bundle model is described as follows. A bundle of $N$ parallel fibers is rigidly clamped at one end and is loaded 
   at the other end. Each individual fiber $i$ has been assigned a breaking threshold $b_i$ of its own, i.e., it can sustain a 
   maximum of $b_i$ stress through it, beyond which it breaks. The breaking thresholds $\{b_i\}$ are drawn from a probability 
   distribution $p(b)$ whose cumulative distribution is $P(b) = \int^{b}_0 p(z)dz$.

      In the fiber bundle model the stress is a conserved quantity. When a fiber breaks, the stress that was acting 
   through it is released and gets distributed among other intact fibers. In the Equal Load Sharing (ELS) version of the fiber 
   bundle model, the released stress is distributed equally among all remaining intact fibers. Using this property the
   relaxation behavior of the bundle can be understood. Two points must be mentioned to describe the model in comparison to 
   the realistic situations: (i) To study the relaxation behavior the bundle is externally loaded by a finite amount of 
   stress per fiber so that a certain fraction of the total number of fibers have their breaking thresholds below the applied load
   and they break immediately. In comparison in standard experiments like `creep test' the breakdown starts from the weakest fiber.
   (ii) In practice fracturing in materials is always associated with the phenomenon of `aging', for example due to 
   thermally-activated environmentally assisted stress corrosion \cite {Ojala2003}. Both these mechanisms have not been incorporated 
   in the fiber bundle model studied here. Moreover, a Local Load Sharing (LLS) version of the fiber bundle model has also
   been studied often in the literature where the released load is distributed to the fibers situated within a local 
   neighborhood of the broken fiber. This version of fiber bundle model is considered to mimic the failure of the actual realistic materials
   more closely.

      In the following we will use $\sigma$ for the 
   notation of the uniform applied load per fiber at the initial stage when all fibers are intact. In comparison $x_t$ will be used 
   to denote the stress per intact fiber after $t$-th relaxation step. Therefore, intially the externally applied load is $F = N\sigma$.
   As a result the bundle relaxes in a series of $T$ successive time steps. The relaxation time $T$ is not really a real time, but 
   it is an integer that represents the number of load redistribution steps for reaching the stable state. At the first step all fibers 
   with breaking thresholds less than $\sigma$ break and therefore each of these fibers releases $\sigma$ amount of stress. Consequently 
   the total amount of stress released is now distributed to $N[1 - P(\sigma)]$ intact fibers on the average, each of them gets the new 
   stress $x_1$ per fiber. Therefore after the first step of relaxation $F = Nx_1[1 - P(\sigma)]$. Similarly the stress per fiber in 
   successive time steps are given by:
\begin {equation}
   F = Nx_1[1 - P(\sigma)] = Nx_2[1 - P(x_1)] = Nx_3[1 - P(x_2)] ..
\end {equation}
   After $T$ steps the system converges to a stable state when the amount of stress released in the last step is no 
   longer sufficient to break even the next fiber in the increasing sequence of breaking thresholds. Therefore on the 
   average $x_{T+1} - x_T < 1/N$. 

%---------------------------------------------------------------------------------
\begin{figure}[t]
\begin {center}
\includegraphics[width=7.0cm]{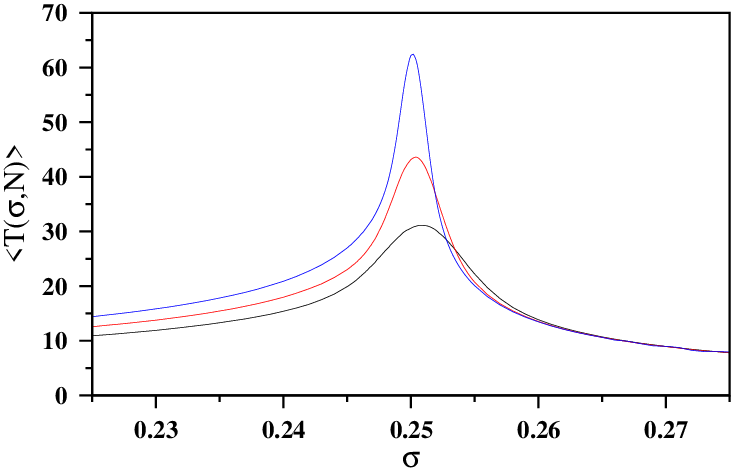}
\end {center}
\caption{(Color online) Plot of $\langle T(\sigma,N) \rangle$ against $\sigma$ for bundle sizes
$N$ = 10000 (black), 30000 (blue) and 100000 (red) with $N$ increasing from bottom to top; $\sigma_c \approx 0.25$.}
\end{figure}
%---------------------------------------------------------------------------------

      In this description when a fiber breaks, it is assumed that the released stress gets distributed instantaneously among all intact fibers
   resulting a bunch of fibers breaking in one relaxation step. In comparison there could be a situation when the stress re-distribution process
   takes place at finite speed \cite {Phoenix1978,Phoenix1979}. The stress acting at all fibers grow uniformly, but the moment the stress 
   reaches the
   breaking threshold of the weakest intact fiber, it breaks. This fiber also releases stress and adds to the rate of growth of stress in each
   fiber. Therefore this model is purely time dependent where real time must pass before failures occur and they occur at the rate of one fiber
   failure at a time. This distinction is important \cite {Phoenix1978}. However, in this paper we consider only the situation where the released 
   stress from a broken fiber is distributed instantaneously.

      In the stable state one writes the applied load $F(x)$ as a function of the stress $x$ 
   per intact fiber at the stable state \cite {Pradhan1,Pradhan2}.
\begin {equation}
   F(x) = Nx[1 - P(x)]. 
\end {equation}
   If for $x = x_c, F(x)$ is maximum then $dF/dx = 0$ yields the following condition:
\begin {equation}
   1 - P(x_c) - x_cp(x_c) = 0. 
\end {equation}
   For a bundle with a uniform distribution of breaking thresholds $p(x) = 1$ one obtains $x_c = 1/2$ and $F_c = N/4$. 
   The total critical applied load $F_c$ corresponds to the critical initial load per fiber \cite {Pradhan1}
\begin {equation}
   \sigma_c = F_c/N = 1/4.
\end {equation}

      Numerically the variation of relaxation times is determined in the following way. We considered a completely intact bundle of $N$ fibers. 
   Uniformly distributed breaking thresholds $\{b_i\}$ were assigned to all fibers. An external load $\sigma$ 
   per fiber was applied to the bundle. The corresponding relaxation time $T(\sigma,N)$ was estimated for this load $\sigma$. This 
   estimation was repeated for different values of $\sigma$ varying from 0 to 1/2 at intervals of $\Delta \sigma = 0.001$ 
   but using the same set of breaking thresholds $\{b_i\}$. The entire calculation was then repeated for a 
   large ensemble of fiber bundles with uncorrelated sets of breaking thresholds $\{b_i\}$ and for different 
   bundle sizes $N$. We observed that in the precritical regime the average relaxation time $\langle T(\sigma,N) \rangle$ increases sharply as 
   $\sigma$ increases and it has a finite but large peak at $\sigma_c \approx 1/4$. The 
   height of the peak increases with increasing $N$ (Fig. 1). In the postcritical regime $\langle T(\sigma,N) \rangle$ gradually 
   decreases as $\sigma$ is increased well beyond $\sigma_c$.

%---------------------------------------------------------------------------------
\begin{figure}[t]
\begin {center}
\includegraphics[width=7.0cm]{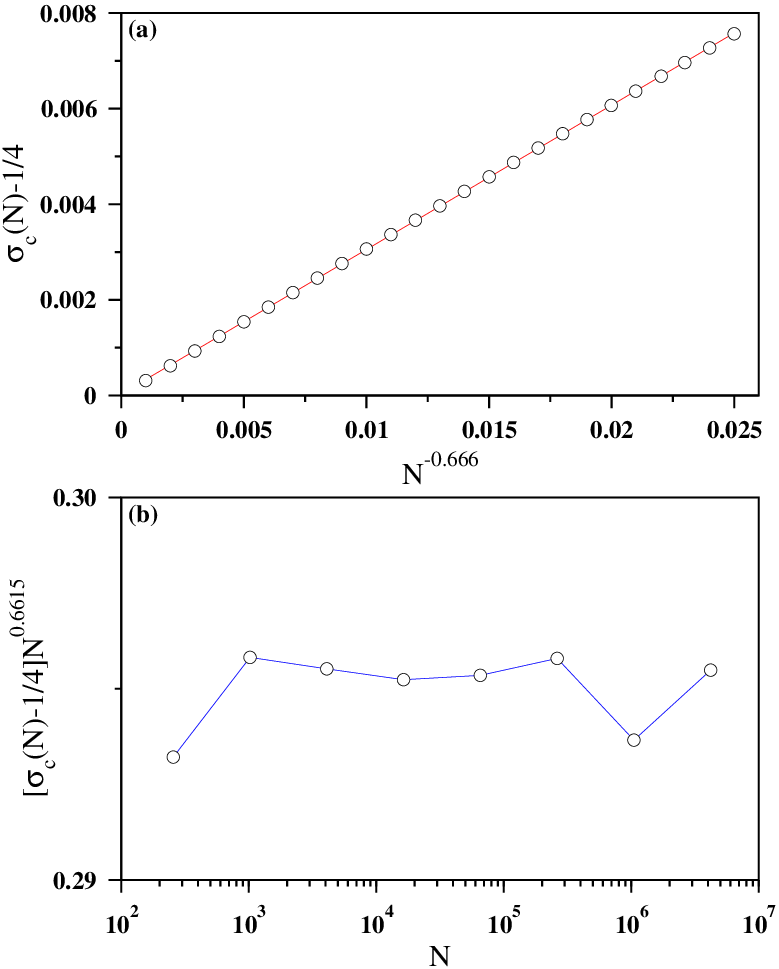}
\end {center}
\caption{(Color online) (a) Plot of $\sigma_c(N)-1/4$ with $N^{-0.666}$ for system sizes up to $N = 31623$ which fits nicely to a straight line
that passes very close to the origin. (b) Data for larger values of $N$ up to $2^{22}$ has been plotted as $[\sigma_c(N)-1/4]N^{0.6615}$ against $\ln (N)$ which exhibits approximately constant variation.}
\end{figure}
%---------------------------------------------------------------------------------

      These numerical results on the relaxation dynamics are supported by mean-field calculations \cite {Pradhan1}. This analysis 
   assumes that for all bundle sizes $N$ the critical threshold $\sigma_c =1/4$ for uniformly distributed breaking thresholds.
   In the vicinity of the critical threshold the variation of the relaxation time with the deviation $|\sigma_c - \sigma|$ has a
   power law form. In the postcritical regime of $\sigma > \sigma_c$
\begin {equation}
T(\sigma,N) \approx \frac{\pi}{2} (\sigma-\sigma_c)^{-1/2}
\end{equation}
   and in the precritical regime of $\sigma < \sigma_c$ and for the range where $(\sigma_c-\sigma) >>1/4N$ \cite {Pradhan1}
\begin {equation}
T(\sigma,N) \approx \frac{\ln (N)}{4} (\sigma_c-\sigma)^{-1/2}.
\end {equation}

      We first noticed that for fiber bundles with uniformly distributed breaking thresholds the average critical applied 
   load per fiber $\sigma_c$ = 1/4 is actually valid only for infinitely large bundles i.e., for $N \to \infty$. Truly, 
   for bundles of finite size the critical load depends on $N$ and we calculated $\sigma_c(N)$ for different bundle sizes
   $N$. We define the critical applied load $\sigma^{\alpha}_c(N)$ for a particular fiber bundle $\alpha$ with a given set of 
   breaking thresholds $\{b_i\}$ as the maximum value of the applied load $\sigma$ per fiber for which the system
   is in the precritical state. This means that if the applied load is increased by the least possible amount to
   include only the next fiber in the increasing sequence of breaking thresholds the system crosses over to the 
   postcritical state. On the average this requires enhancing the applied load by $1/N$.

%---------------------------------------------------------------------------------
\begin{figure}[t]
\begin {center}
\includegraphics[width=7.0cm]{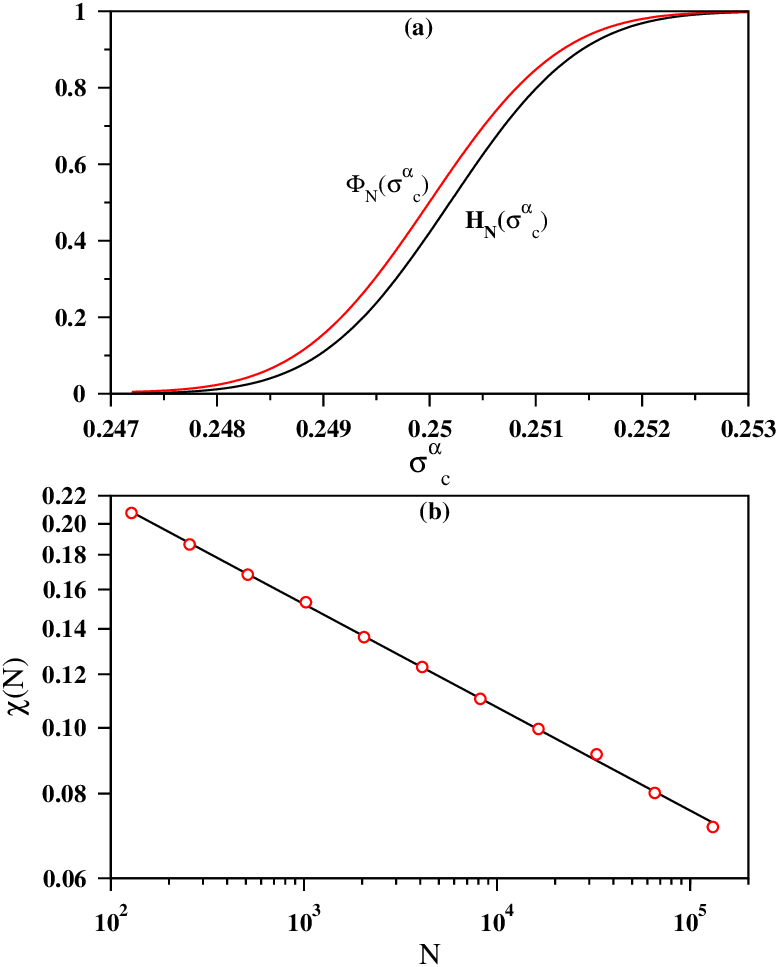}
\end {center}
\caption{(Color online) (a) Plot of the cumulative probability distribution of ${\cal H}_N(\sigma^{\alpha}_c)$ for $N = 2^{16}$ and for
a sample size of $10^6$ bundles with red color. The cumulative distribution of the Gaussian approximation $\Phi_N(\sigma^{\alpha}_c)$ 
has also been
plotted using black color. (b) The maximal difference $\chi(N)$ between two cumulative distributions has been plotted against $N$ using 
the log - log scale. The slope is found to be 0.155(5).}
\end{figure}
%---------------------------------------------------------------------------------

      The value of $\sigma^{\alpha}_c(N)$ is numerically determined using the
   bisection method. The simulation starts with a pair of guessed values for $\sigma^{\alpha}_{pre}$ and 
   $\sigma^{\alpha}_{post}$ corresponding to the precritical and postcritical states respectively. In the precritical state 
   the relaxation dynamics stops without breaking the entire bundle whereas in the postcritical state all fibers in the 
   bundle break. The bundle is then subjected to the mean of two stress values, $\sigma = (\sigma^{\alpha}_{pre} + \sigma^{\alpha}_{post})/2$
   and then relaxed. If the final stable state is precritical $\sigma^{\alpha}_{pre}$ is raised to $\sigma$ otherwise $\sigma^{\alpha}_{post}$ 
   is reduced to 
   $\sigma$. This procedure is terminated when $\sigma^{\alpha}_{post} - \sigma^{\alpha}_{pre} \le 1/N$ and at this stage we define 
   $\sigma^{\alpha}_c(N) = (\sigma^{\alpha}_{post} + \sigma^{\alpha}_{pre})/2$. This iteration is repeated for a large number of un-correlated bundles $\alpha$ and their 
   critical loads are averaged to obtain $\sigma_c(N) = \langle \sigma^{\alpha}_c(N) \rangle$ for a fixed bundle size $N$. Next the entire 
   calculation has been repeated for different values of $N$.

      There exists a more straight forward way to calculate the initial critical load per fiber $\sigma^{\alpha}_c(N)$ of a specific 
    fiber bundle. If 
    $b^{\alpha}_{(1)}$, $b^{\alpha}_{(2)}$, $b^{\alpha}_{(3)}$, ... , $b^{\alpha}_{(N)}$ are the breaking thresholds ordered in an 
    increasing sequence, then 
\begin {equation}
\sigma^{\alpha}_c(N) = \verb+max + \Bigg\{ b^{\alpha}_{(1)}, \frac {N-1}{N} b^{\alpha}_{(2)}, \frac {N-2}{N}b^{\alpha}_{(3)}, ... , \frac{1}{N}b^{\alpha}_{(N)} \Bigg\}.
\end {equation}
    Both methods need to order the breaking thresholds only once in increasing sequence and this makes the major share of the CPU.
    The well known Quicksort method takes CPU of the order of $N \ln N$ \cite {Quicksort}. Comparing the two methods the bisection 
    method takes little more time, e.g., for a single bundle of $N = 2^{24}$ the bisection
    method takes $\approx$ 1.15 times the time required in the second method. 

%---------------------------------------------------------------------------------
\begin{figure}[t]
\begin {center}
\includegraphics[width=8.0cm]{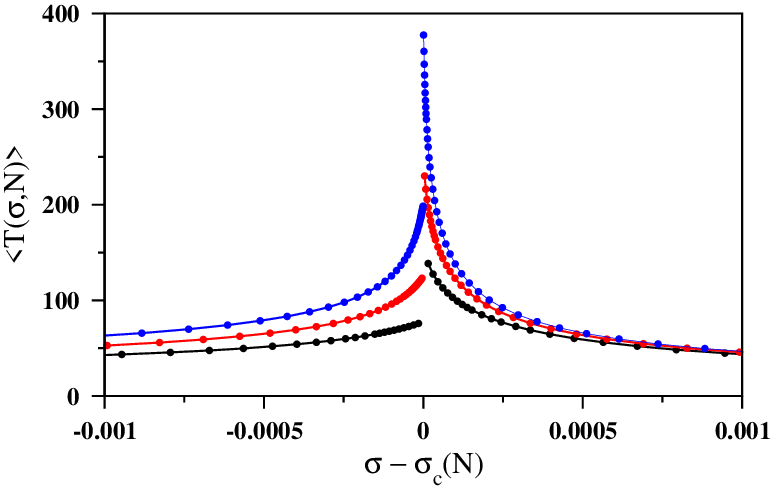}
\end {center}
\caption{(Color online) The average relaxation time $\langle T(\sigma,N) \rangle$ has been plotted with the deviation from the critical point 
$\sigma - \sigma_c(N)$ per fiber for $N = 2^{16}$ (black), $2^{18}$ (red) and $2^{20}$ (blue) with $N$ increasing from bottom to top. 
More specifically for each fiber bundle $\alpha$ first its 
critical point $\sigma^{\alpha}_c$ is determined. Then for the same bundle the relaxation times are measured for different 
deviations $\Delta \sigma = \sigma - \sigma^{\alpha}_c$ and then averaged over many different bundles.}
\end{figure}
%---------------------------------------------------------------------------------

      We assume that the average values of the critical load per fiber $\sigma_c(N)$ for the bundle size $N$ converges to a specific 
   value $\sigma_c = \sigma_c(\infty)$ as $N \to \infty$ according to the following form:
\begin {equation}
\sigma_c(N)-\sigma_c = AN^{-1/\nu}
\end {equation}
   where $\nu$ is a critical exponent. Accordingly $\sigma_c(N)$ values have been plotted in Fig. 2. A plot of $\sigma_c(N)-\sigma_c$ 
   against $N^{-1/\nu}$ using $\sigma_c=0.25$ and $1/\nu = 0.666$ fits to an excellent straight line. The 
   least square fitted straight line misses the origin very closely and has the form $\sigma_c(N)-1/4$ = 3.33 $\times 10^{-5}$ + 0.302$N^{-1/\nu}$.
   In Fig. 2(b) data for larger values of $N$ have been
   plotted as $[\sigma_c(N)-1/4]N^{0.6615}$ against $N$ on a lin - log scale. The intermediate part appears approximately constant
   implying again that $1/\nu \approx 0.662$. Our conclusion is $\nu = 1.50(2)$ and $\sigma_c=0.2500(1)$.
   We conjecture that the finite size correction exponent is $\nu=3/2$ and $\sigma_c=1/4$ exactly \cite {Daniels}.

%---------------------------------------------------------------------------------
\begin{figure}[top]
\begin {center}
\includegraphics[width=7.0cm]{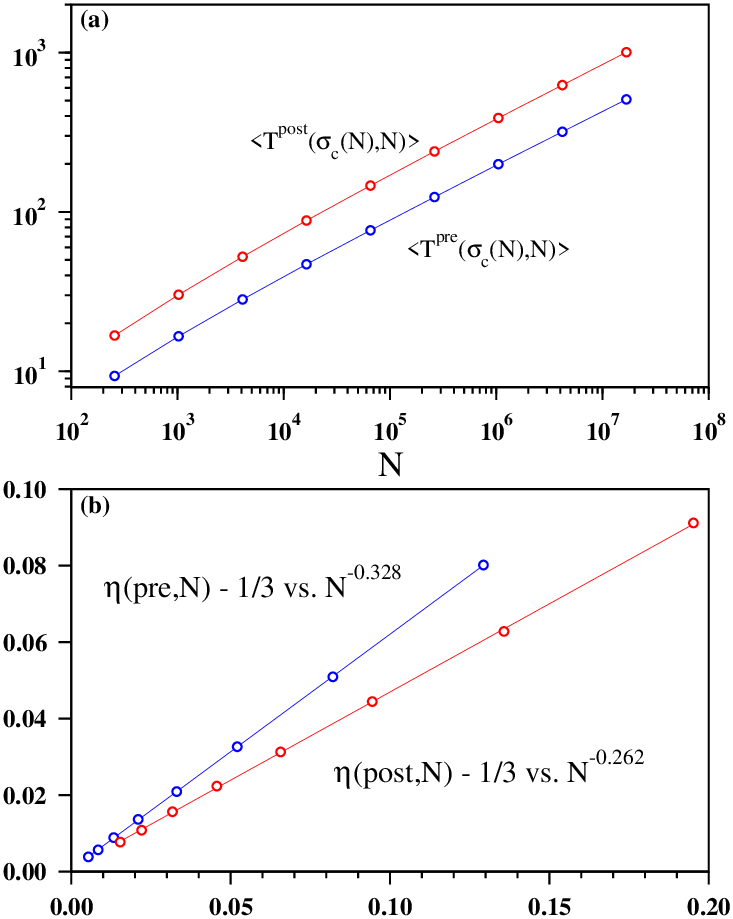}
\end {center}
\caption{(Color online) (a) Plots of the average maximal relaxation time $\langle T^{post}(\sigma_c(N),N) \rangle$ in the postcritical 
regime (red) and the average maximal relaxation time $\langle T^{pre}(\sigma_c(N),N) \rangle$
in the precritical regime (blue) against the system size $N$ using log - log scale. Both plots exhibit
certain amount of curvature. (b) Slopes $\eta(N)$ between successive points in (a) are estimated and $\eta(N) - 1/3$ are extrapolated 
against $N^{-0.328}$ and $N^{-0.262}$. The solid lines are obtained by least square fits whose 
intercepts are 0.00061 and 0.00085 for the precritical and postcritical regimes respectively.}
\end{figure}
%---------------------------------------------------------------------------------

 %---------------------------------------------------------------------------------
\begin{figure}[t]
\begin {center}
\includegraphics[width=7.0cm]{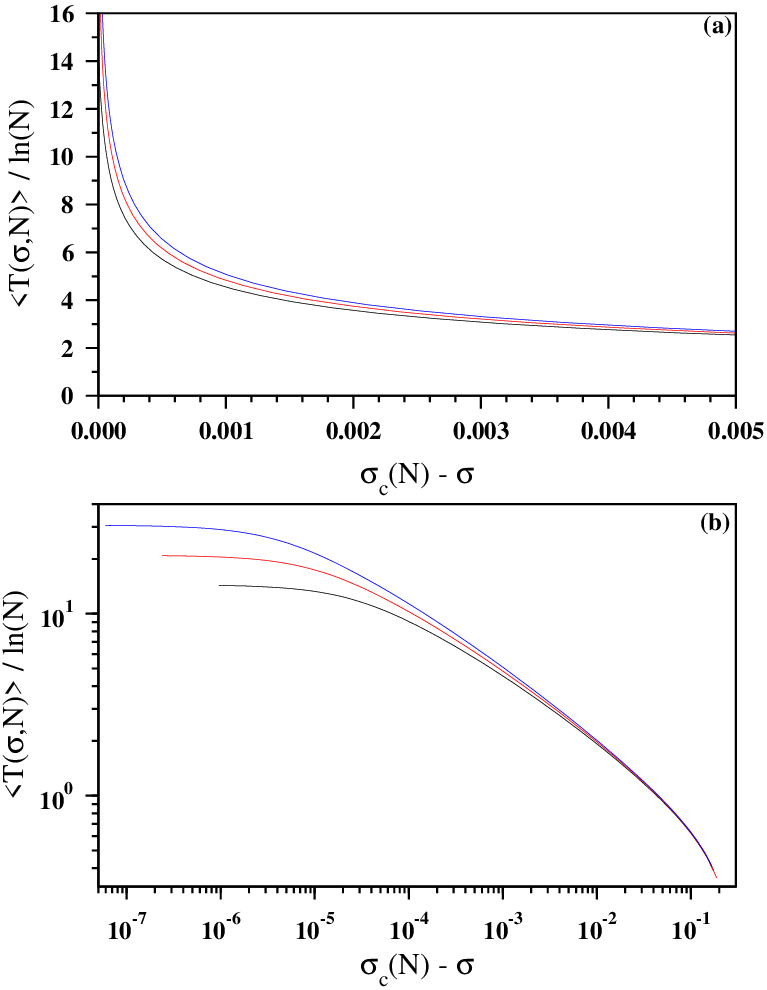}
\end {center}
\caption{(Color online) 
Comparison with the similar plots in \cite {Pradhan1}.
Plot of $\langle T(\sigma,N) \rangle / ln (N)$ against $\Delta \sigma = \sigma_c(N)-\sigma$ for
the precritical regime but for much smaller window of $\Delta \sigma$ = 0.005 and 
for $N = 2^{20}$ (black), $2^{22}$ (red) and $2^{24}$ (blue) with $N$ increasing from left to right. 
(a) On a lin - lin scale the three plots get separated from one another as $\Delta \sigma \to 0$.
(b) The data in (a) has been replotted on a log - log scale and the absence of data collapse
is more distinctly visible in this plot, with $N$ increasing from bottom to top.
}
\end{figure}
%---------------------------------------------------------------------------------

      These results are known in the literature from analytical studies \cite {Smith1982,McCartney1983}. 
      It has been estimated that \cite {Smith1982}
\begin {equation}
\sigma_c(N) = \sigma_c+ 0.996N^{-2/3}\beta_c
\end {equation}
   where, 
\begin {equation}
\beta_c = \Bigg[\frac{P'(x_c)^2x_c^4}{2P'(x_c)+x_cP''(x_c)}\Bigg]^{1/3}
\end {equation}
   where $P'(x) = dP(x)/dx = p(x)$. In our case with the uniformly distributed breaking thresholds in the range $\{0,1\}$;
   $P(x) = x$ which gives $\sigma_c = 1/4$, $x_c = 1/2$, $P'(x) =1$ and $P''(x) = 0$ for all $0 < x < 1$ which makes $\beta_c = (1/2)^{5/3}
   \approx 0.3150$. This gives
\begin {equation}
\sigma_c(N) - \sigma_c = 0.996 N^{-2/3} \beta_c = 0.3137N^{-2/3}.
\end {equation}
   Therefore apart from the exponent $\nu = 3/2$ one can also check the value of the amplitude $A$ which is estimated numerically as 0.302
   compared to its analytically obtained value of 0.3137. The correspondence is quite good and this is a confirmation of the rigorous
   result of \cite {Smith1982}.

      For a large uncorrelated sample of 
   fiber bundles of a specific size $N$ the critical loads per fiber $\sigma^{\alpha}_c(N)$ is known to have a Gaussian 
   distribution around its mean value $\sigma_c(N) = \langle \sigma^{\alpha}_c(N) \rangle$. Let its cumulative distribution be 
   denoted by ${\cal H}_N(\sigma_c^{\alpha})$. As the bundle size increases to very large values this cumulative distribution 
   approaches to its Gaussian approximation $\Phi_N(\sigma^{\alpha}_c)$ which is also the cumulative distribution of the Gaussian form: 
\begin {equation}
A\exp\{-(\sigma^{\alpha}_c-\sigma_c)^2/(2s^2)\}
\end {equation}
   where $\sigma_c=x_c(1-P(x_c)) = 1/4$, 
   $s=\gamma_c N^{-1/2}$ and $\gamma_c = x_c\{P(x_c)(1-P(x_c))\}^{1/2}$. Using these results it has been shown that \cite {Smith1982}
\begin {equation}
\chi(N) = \verb+max + \big\lvert {\cal H}_N(\sigma^{\alpha}_c) - \Phi_N(\sigma^{\alpha}_c) \big\rvert < KN^{-1/6}.
\end {equation}
   This relation has also been verified numerically in Fig. 3(a). For the bundle size $N = 2^{16}$, the cumulative distribution 
   ${\cal H}_N(\sigma^{\alpha}_c)$ obtained from simulation and the $\Phi_N(\sigma^{\alpha}_c)$ obtained from the Gaussian approximation have been plotted.
   In simulation, a sample size of $10^6$ bundles have been studied for each bundle size $N$. These critical loads $\sigma^{\alpha}_c$s 
   have been arranged in the
   increasing order, so that the number of such thresholds below a certain $\sigma^{\alpha}_c$ is simply the ${\cal H}_N(\sigma^{\alpha}_c)$. 
   For each
   of these $\sigma^{\alpha}_c$ values the cumulative Gaussian function $\Phi_N(\sigma^{\alpha}_c)$ has been calculated. The absolute value of the difference
   between these two distributions have been estimated for each $\sigma^{\alpha}_c$ and their maximal value $\chi(N)$ has been found out.
   In Fig. 3(b) the function $\chi(N)$ has been plotted with $N$ on a log - log scale for eleven different bundle sizes. A power law variation
   of $\chi(N)$ has been observed:
\begin {equation}
\chi(N) \sim N^{-\kappa}
\end {equation}
   with $\kappa = 0.155(5)$ (Fig. 3(b)).

 %---------------------------------------------------------------------------------
\begin{figure}[t]
\begin {center}
\includegraphics[width=7.0cm]{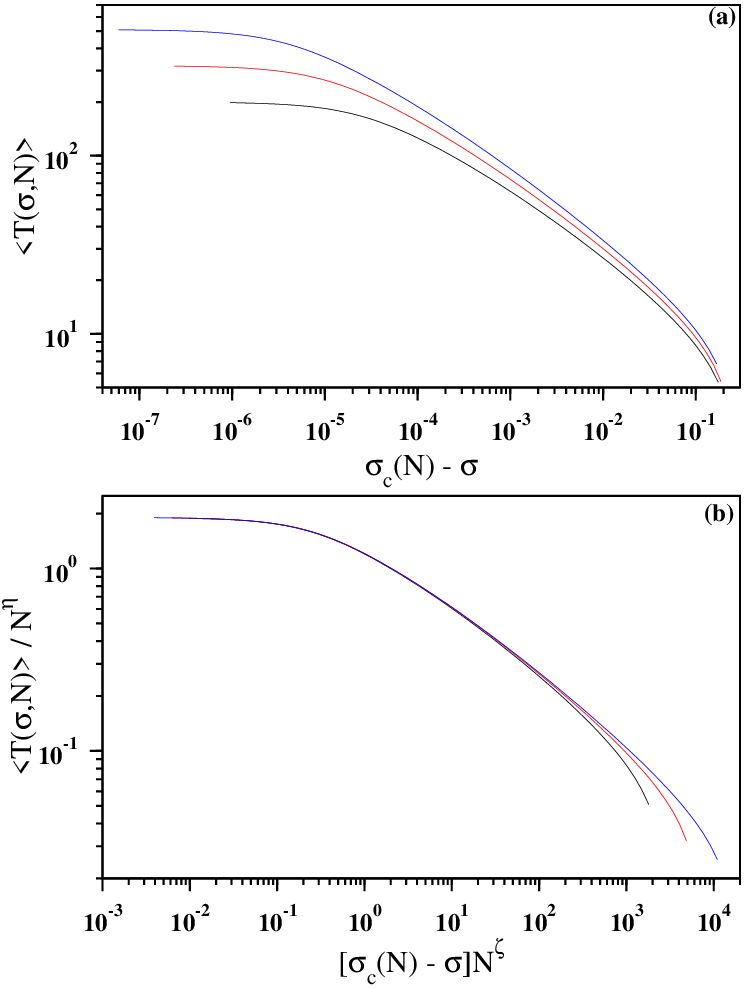}
\end {center}
\caption{(Color online) Scaling for the precritical regime.
(a) Plot of $\langle T(\sigma,N) \rangle$ against $\sigma_c(N)-\sigma$
    and for $N = 2^{20}$ (black), $2^{22}$ (red) and $2^{24}$ (blue) with $N$ increasing from bottom to top. 
(b) The data in (a) has been scaled suitably:
$\langle T(\sigma,N) \rangle / N^{\eta}$ against $[\sigma_c(N) - \sigma]N^{\zeta}$
exhibits a good collapse of the data as $\Delta \sigma \to 0$ with $\eta = 0.336$ and $\zeta=0.666$.
Here $N$ increases from left to right.
}
\end{figure}
%---------------------------------------------------------------------------------

      Once we know the system size dependent critical loads $\sigma_c(N)$ we studied how the average relaxation time 
   $\langle T(\sigma,N) \rangle$ diverges as the critical load is approached. For every bundle $\alpha$ we first calculated its critical 
   load $\sigma^{\alpha}_c$ using the bisection method as described above. Then for the same bundle $\alpha$ we calculated the relaxation 
   times for 
   certain pre-fixed deviations $|\Delta \sigma| = |\sigma^{\alpha}_c - \sigma|$ from the critical stress
   and then averaged over different uncorrelated bundles. Fig. 4 shows how $\langle T(\sigma,N) \rangle$ approaches the critical relaxation
   time as $\sigma \to \sigma_c(N)$. We observe that the limiting relaxation times 
   as $|\Delta \sigma| \to 0$ for the precritical and postcritical states are distinctly different and call them as
   $\langle T^{pre}(\sigma_c(N),N) \rangle$ and $\langle T^{post}(\sigma_c(N),N) \rangle$ respectively.

      Next we calculated the average relaxation times when the applied load per fiber takes the critical load. 
   For each bundle $\alpha$ we calculated two values of $T$: $T^{pre}$ denotes the largest value of $T$ in the precritical 
   state and $T^{post}$ is the largest value of $T$ in the postcritical state. We see that $T^{post}$ is much larger than 
   $T^{pre}$ and when averaged over a large sample size $\langle T^{post} \rangle / \langle T^{pre} \rangle$ approaches to 2
   as $N \rightarrow \infty$.

 %---------------------------------------------------------------------------------
\begin{figure}[t]
\begin {center}
\includegraphics[width=7.0cm]{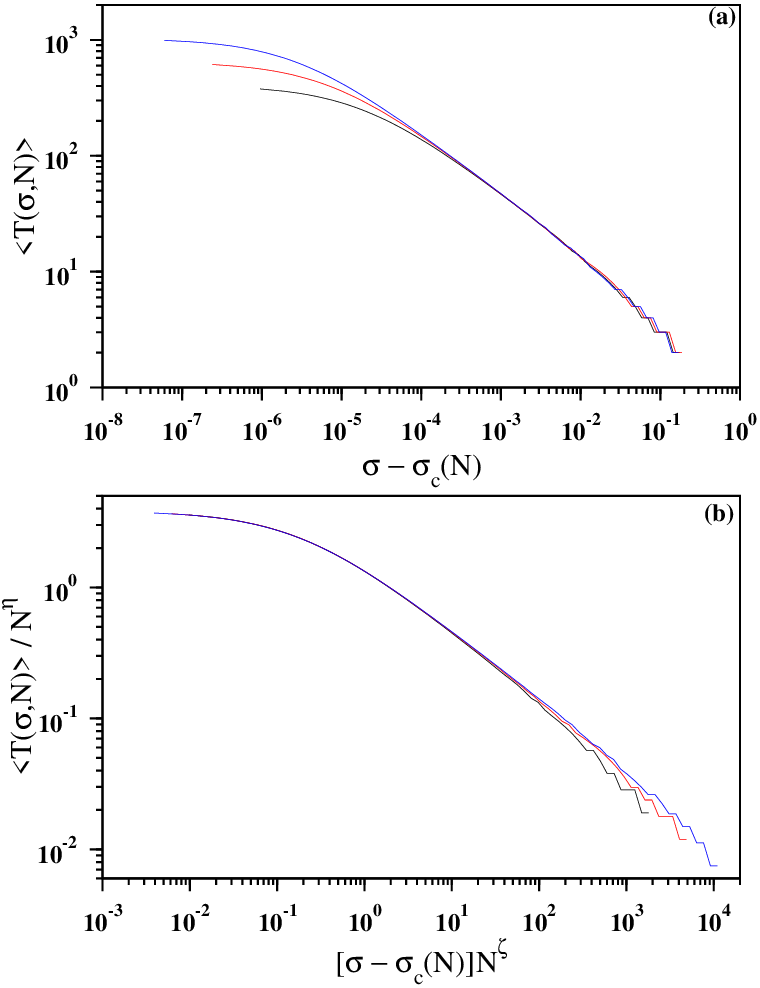}
\end {center}
\caption{(Color online) Scaling for the postcritical regime.
(a) Plot of $\langle T(\sigma,N) \rangle$ against $\sigma - \sigma_c(N)$
    and for $N = 2^{20}$ (black), $2^{22}$ (red) and $2^{24}$ (blue) with $N$ increasing from bottom to top. 
(b) The data in (a) has been scaled suitably:
$\langle T(\sigma,N) \rangle / N^{\eta}$ against $[\sigma - \sigma_c(N)]N^{\zeta}$
exhibits a good collapse of the data as $\Delta \sigma \to 0$ with $\eta = 0.336$ and $\zeta=0.666$.
Here $N$ increases from left to right.
}
\end{figure}
%---------------------------------------------------------------------------------

      In Fig. 5(a) we plot $\langle T^{pre}(\sigma_c(N),N) \rangle$ and $\langle T^{post}(\sigma_c(N),N) \rangle$ against $N$ on a 
   log - log scale for a wide range of values of $N$ extending from $2^8$ to $2^{24}$, at each step the system size being 
   increased by a factor of 4.
   Both curves are nearly straight and parallel for large $N$ but have slight curvature for small $N$. Upto $N=2^{22}$ the averaging has been done for $10^6$
   independent configurations and for $N=2^{24}$ a total of 409000 independent configurations have been used. Therefore the data points are 
   accurate enough to be analyzed
   more precisely. We define the slope between successive points in Fig. 5(a) as $\eta(N)$ and observe that these slopes gradually approach
   1/3 for both plots. We estimated suitable extrapolation methods minimizing the errors and in Fig. 5(b) extrapolated
   $\eta(pre,N)-1/3$ against $N^{-0.328}$ and $\eta(post,N)-1/3$ against $N^{-0.262}$ for the precritical and postcritical states respectively.
   Individual plots fit excellent to straight lines and their intercepts with the vertical axes are 0.00061 and 0.00085 respectively.
   We conclude that when the system is loaded with the precise value of the critical stress the relaxation time
   grows as a power of the system size as: 
\begin {equation}
   \langle T(\sigma_c(N),N) \rangle \sim N^{\eta},
\end {equation}
   with $\eta=0.333(1)$.

%---------------------------------------------------------------------------------
\begin{figure}[t]
\begin {center}
\includegraphics[width=7.0cm]{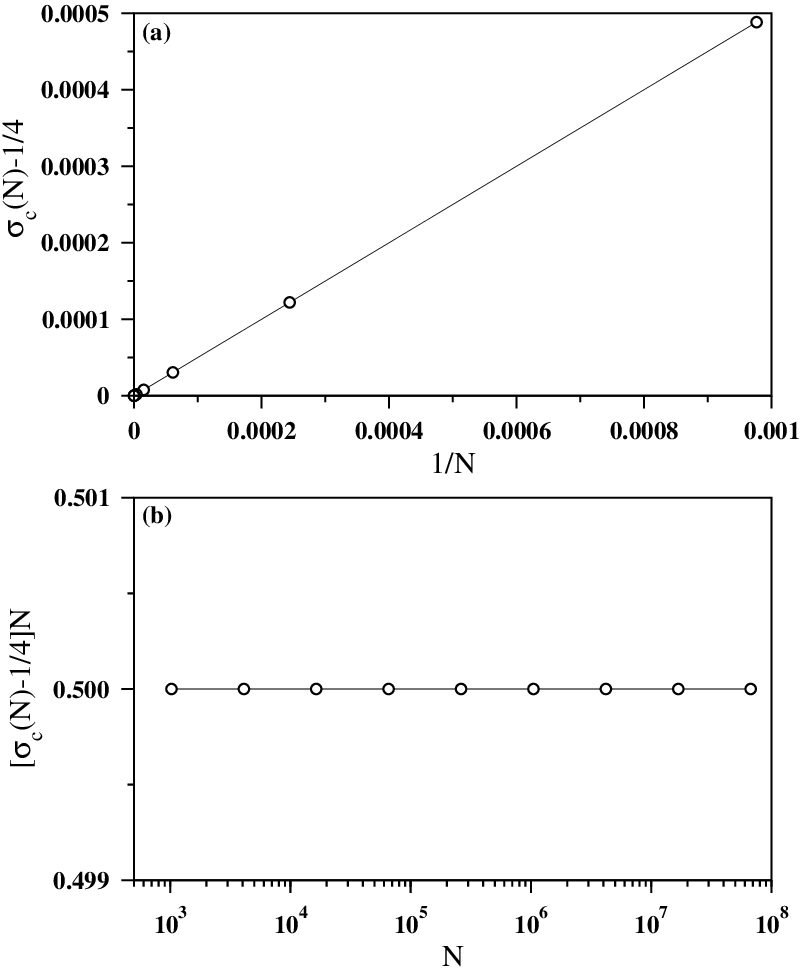}
\end {center}
\caption{(Color online) The variation of the critical load $\sigma_c(N)$ on the system size $N$ in the deterministic case.
(a) Plot of $\sigma_c(N) - 1/4$ vs. $1/N$ gives an excellent straight line that passes very close to the origin:
$\sigma_c(N)-1/4 = -1.3 \times 10^{-15}+0.5/N$.
(b) Same data as in (a) but here $(\sigma_c(N) - 1/4)N$ has been plotted with $N$ on a semi-log scale
and the plot exhibits a horizontal straight line indicating that quite possibly $\sigma_c(N) = 1/4 + \frac{1}{2N}$.
}
\end{figure}
%---------------------------------------------------------------------------------

      Our data for relaxation times away from the critical point are compared in Fig. 6 with the similar data presented in 
   \cite {Pradhan1} which assumed $\sigma_c = 1/4$ for all bundle sizes $N$.
   In Fig. 6(a) $\langle T(\sigma,N) \rangle / \ln (N)$ has been plotted against $\sigma_c(N) - \sigma$. The large sample sizes 
   yielded data points with very little noise and allowed us to plot for much smaller window size, i.e.,
   $\Delta \sigma = 0.005$ compared to 0.05 in \cite {Pradhan1}. It is observed that three curves separate out distinctly and
   systematically from one another as $\Delta \sigma \to 0$. The same data have been plotted in Fig. 6(b) using a log - log scale.
   In this figure the absence of data collapse is even more pronounced. We explain the difference in the following way.
   The claimed validity of data collapse exhibited in \cite {Pradhan1} is for a window size 10 times larger than ours. When we
   reduced the window size and thus approached the critical point even closer, the scaling by $\ln (N)$ no longer works. 
   We see below that instead a simple power law scaling works quite well.

%---------------------------------------------------------------------------------
\begin{figure}[t]
\begin {center}
\includegraphics[width=7.0cm]{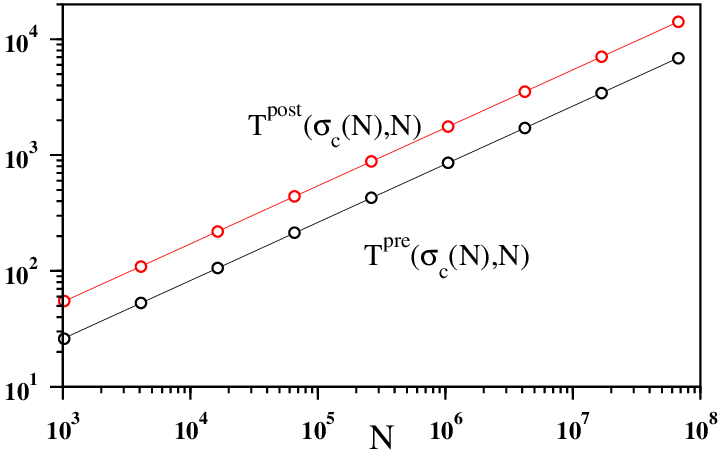}
\end {center}
\caption{(Color online) The deterministic case where breaking thresholds for individual fibers are uniformly spaced at an interval of 
$1/N$. The average relaxation time $T(\sigma_c(N),N)$ has been plotted with the bundle size $N$ on a 
log - log scale for $N = 2^{10}$ to $2^{26}$. The slopes are 0.502 and 0.501 for the precritical and postcritical regimes respectively.
}
\end{figure}
%---------------------------------------------------------------------------------

      This data $\langle T(\sigma,N) \rangle$ against $\sigma_c(N) - \sigma$ for the precritical regime have been replotted
   in Fig. 7(a). The plots for the three $N$ values are completely separated. Now a finite size scaling of the two axes have 
   been done in Fig. 7(b) by appropriate powers of the bundle size $N$. This indeed results an excellent collapse of the data 
   for the three different bundle sizes. This implies that the following scaling form may describe the collapse:
\begin {equation}
\langle T(\sigma,N) \rangle / N^{\eta} \sim {\cal G}[\{\sigma_c(N)-\sigma\}N^{\zeta}]
\end {equation}
   where ${\cal G}(y)$ is an universal scaling function of the scaled variable $y$.
   The best possible tuned values of the scaling exponents obtained are $\eta = 0.336$ and $\zeta = 0.666$.
   The collapsed plots have two different regimes, an initial constant part for very small values of $\Delta \sigma = \sigma_c(N)-\sigma$.
   In this regime the scaled variable $\langle T(\sigma,N) \rangle / N^{\eta}$ is a constant, say ${\cal C}$. This means
   $\langle T(\sigma,N) \rangle = {\cal C}N^{\eta}$ which is the retrieval of the Eqn. (15). 
   Again the constant regime of $\langle T(\sigma,N) \rangle / N^{\eta}$ is extended approximately up to 
   $\{\sigma_c(N)-\sigma\}N^{\zeta} \approx 1$. This implies that the width of the constant regime is:
\begin {equation}
\sigma_c(N) - \sigma \sim N^{-\zeta}.
\end {equation}
   The exponent $\zeta$ can also be interpreted in the following way. For a certain bundle size $N$ there exists a specific
   value of $|\Delta \sigma(eq,N)|$ where $\langle T(pre,\sigma,N) \rangle = \langle T(post,\sigma,N) \rangle$. Around this window size
   $\langle T(pre,\sigma,N) \rangle > \langle T(post,\sigma,N) \rangle$ for $|\Delta \sigma(N)| > |\Delta \sigma(eq,N)|$ and
   $\langle T(pre,\sigma,N) \rangle < \langle T(post,\sigma,N) \rangle$ for $|\Delta \sigma(N)| < |\Delta \sigma(eq,N)|$. We
   have verified that $|\Delta \sigma(eq,N)|$ also approaches to zero as $N^{-\zeta}$ with $\zeta \approx 0.666$. 
   The exponent $\zeta$ is recognized as the inverse of the exponent $\nu$ defined in Eqn. (8).

%---------------------------------------------------------------------------------
\begin{figure}[t]
\begin {center}
\includegraphics[width=7.0cm]{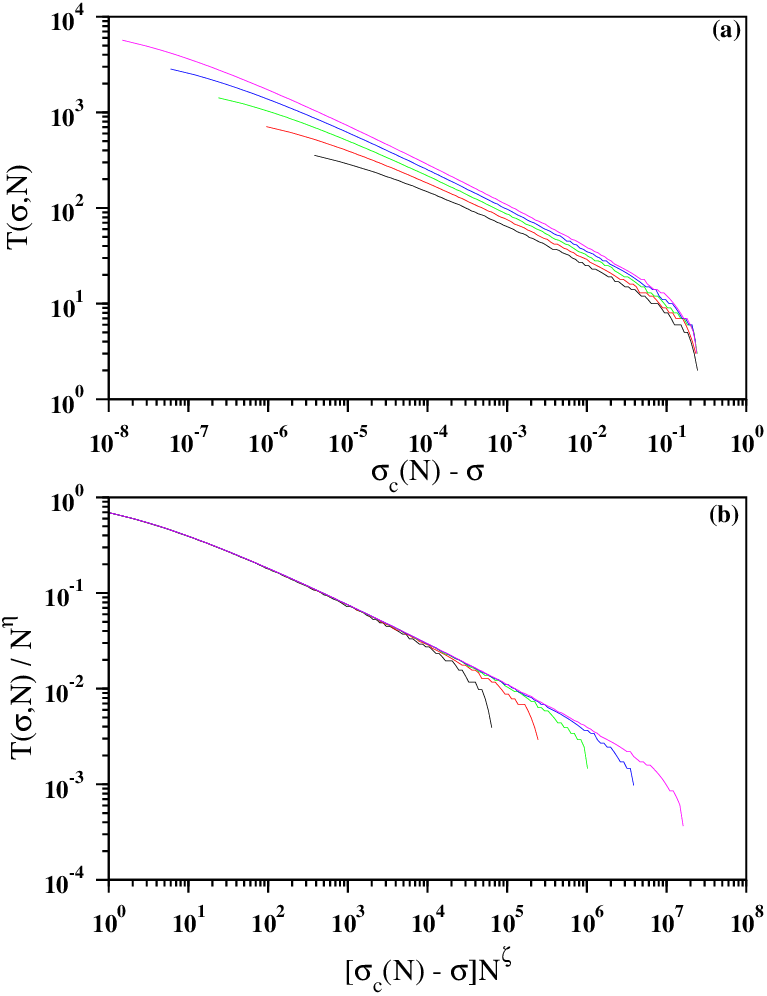}
\end {center}
\caption{(Color online) The deterministic case: 
(a) Plot of $T(\sigma,N)$ against ($\sigma_c(N)-\sigma$) for $N$ = $2^{18}$ to $2^{26}$, the bundle size is increased by a factor
of 4 at each step with $N$ increasing from bottom to top.
(b) A finite size scaling analysis of the data in (a) using the scaling form in Eqn. (16) with $\eta=1/2$ and $\zeta=1$.
Here $N$ increases from left to right.
}
\end{figure}
%---------------------------------------------------------------------------------

      Beyond this constant regime is the power law regime. Assuming that the scaling in Fig. 7(b) is valid for all bundle sizes till 
   $N \to \infty$ one would expect that an $N$ independent power law form holds in this limit:
\begin {equation}
\langle T(\sigma) \rangle \sim (\sigma_c-\sigma)^{-\tau}
\end {equation}
   To ensure that Eqn. (18) indeed holds good we need to assume ${\cal G}(y) \sim y^{-\tau}$ which implies the following scaling relation:
\begin {equation}
-\tau \zeta+\eta = 0
\end {equation}
   and therefore $\tau=\eta/\zeta = 0.50(1)$.

      Similar plots for the postcritical regime have been shown in Fig. 8. In Fig. 8(a) 
   $\langle T(\sigma,N) \rangle$ has been plotted with $\sigma - \sigma_c(N)$ using a log - log scale.
   The scaling of the same data have been shown in Fig. 8(b) as $\langle T(\sigma,N) \rangle / N^{\eta}$ against
   $[\sigma - \sigma_c(N)]N^{\zeta}$ which again show nice data collapse. Here also we obtained very similar values
   of $\eta = 0.336$ and $\zeta=0.666$. The range of validity of the finite size scaling form in Eqn. (16) may be determined
   from Fig. 7(b). Here the data collapse is observed from the smallest value of $[\sigma_c(N)-\sigma]N^{\zeta}$
   to about 100. Therefore the range of validity is $1/N < (\sigma_c(N)-\sigma) < 100N^{-\zeta}$. 

      The entire set of calculations have been repeated with breaking thresholds for fibers drawn from the Weibull
   distributions $P(\sigma) = 1 - \exp(-\sigma^{\rho})$ with the shape parameter $\rho=5$ and the scale parameter 1. 
   A similar use of Smith's results yield
   $\sigma_c = (\rho e)^{-1/\rho}$, $x_c = \rho^{-1/\rho}$ and $\beta_c = \rho^{-(\rho+3)/(3\rho)}e^{-1/(3\rho)}$. Using
   $\rho = 5$ gives $\beta_c = 5^{-8/15}e^{-1/15} = 0.3965$. This gives
\begin {equation}
\sigma_c(N) - \sigma_c = 0.3949 N^{-2/3}.
\end {equation}
   We have estimated the values of $\sigma_c(N)$ numerically for five different bundle sizes: $2^{16}$ to $2^{24}$ increased
   by a factor of 4 at every step. Plotting them against $N^{-2/3}$ and on extrapolation as $N \to \infty$ we have obtained
   $\sigma_c(\infty) = 0.5934(10)$ and $A = 0.392(4)$ which are very much consistent with the analytical results.
   Further we have estimated the exponents $\nu, \eta, \zeta$ and $\tau$
   which are also quite consistent with similar exponents with uniformly distributed breaking thresholds. The critical 
   points as well as the critical exponents are summarized in Table I.

      A simpler version of the fiber bundle model is the deterministic case where the breaking thresholds of the
   $N$ fibers are uniformly spaced as $b_i=n/N$ where $n=1,2,3, ... ,N$ \cite {Pradhan2002}. For this deterministic case no
   averaging is necessary and therefore studying only one configuration is sufficient. The breaking thresholds are already
   in the increasing order. In spite of the absence of randomness the system has a very systematic dependence on
   the size of the bundle $N$.

%--------------------------------------------------------------------------------------------------------------------------
\begin{table}[t]
\begin {tabular}{crrrrr} \\ \hline \vspace*{0.1cm}
P($\sigma$) & $\sigma_c$ & $\nu$   & $\eta$ & $\zeta$ & $\tau$ \\ \hline \hline
Uniform     & 0.250(1)   & 1.50(1) & 0.336(5) & 0.666(5) & 0.50(1)  \\
$P(x)=x$    & 1/4        & 3/2     & 1/3      & 2/3      & 1/2      \\ \hline 
Weibull            & 0.593(1)      & 1.50(1) & 0.335(5) & 0.663(5) & 0.50(1)  \\
$P(x)=1-e^{-x^5}$ & $(5e)^{-1/5}$ &    3/2  & 1/3      & 2/3      & 1/2 \\ \hline
DFBM               & 0.2500(1)     &    1.00(1) & 0.50(1)  & 1.00(1)  & 0.50(1) \\
                   & 1/4           &         1  &       1/2&   1      &  1/2  \\ \hline 
\end {tabular}
\caption{Summary of the values of critical points and critical exponents for different distributions of breaking
thresholds, uniform and Weibull. The results for the deterministic fiber bundle model (DFBM) are also included. For each
distribution the numerical estimates are given in the first row and the conjectured values are given in the second row.}
\end {table}
%--------------------------------------------------------------------------------------------------------------------------

      In Fig. 9(a) we show the plot of $\sigma_c(N)-1/4$ with $1/N$ for different values of $N$ starting from $2^{10}$ to $2^{26}$
   and we see that all points fall on a straight line. By a least square fit it is seen that these points fit excellently
   to a straight line passing very close to the origin: $\sigma_c(N)-1/4 = -1.3 \times 10^{-15}+0.5/N$. To see the
   variation even more distinctly we plot in Fig. 9(b) $[\sigma_c(N)-1/4]N$ against $N$ on a lin - log scale. 
   The fitted straight line is very much parallel to the $\ln (N)$ axis and has the value 0.5000(1). We conjecture that the
   exact form of variation may be  $\sigma_c(N)-1/4 = \frac{1}{2N}$.

      The maximal relaxation times $T^{pre}(\sigma_c(N),N)$ and $T^{post}(\sigma_c(N),N)$ at the critical loads have also been 
   calculated for the deterministic
   fiber bundle model. We show both these plots in Fig. 10 against $N$ using a log - log scale for the same sizes of the
   fiber bundles as in Fig. 9. Unlike the stochastic fiber bundles here the plots fit nicely to straight lines without
   any systematic curvatures for small bundles. From slopes we estimate the exponents as 0.502 and 0.501 respectively
   for the precritical and postcritical regimes. We conclude a common value of $\eta=\eta_{pre}=\eta_{post}=0.500(5)$ for both
   exponents.

      Finally a finite size scaling of the relaxation times as a function of deviation from the critical load has also
   been exhibited in Fig. 11. In Fig. 11(a) $T(\sigma_c(N),N)$ has been plotted against $\sigma_c(N)-\sigma$ using again the 
   double logarithmic scales for bundles of sizes $N$ = $2^{18}$ to $2^{26}$. It is observed that each curve has considerable 
   curvature, yet it is apparent that as the bundle sizes become increasingly larger they tend to assume a power law form.
   We again tried a finite size
   scaling of these in Fig. 11(b) and tried if a data collapse for very small deviations from the critical point is possible. 
   Plotting $T(\sigma,N)/N^{1/2}$ against $(\sigma_c(N)-\sigma)N$ we do find a reasonably good collapse for the small values
   of $(\sigma_c(N)-\sigma)N$. From the scaling exponent values $\eta=1/2$ and $\zeta=1$ we conclude a value for the exponent 
   $\tau=1/2$ for the precritical regime. The same exponent values for $\eta, \zeta$ and $\tau$ are also concluded for the 
   postcritical regime.

      To summarize we have revisited the relaxation behavior of the fiber bundle model with equal load sharing dynamics using 
   extensive numerical calculations. Numerical values of a number of critical points and exponents have been estimated very accurately
   and have been compared with their analytical counterparts known in the literature. For breaking thresholds distributed uniformly and 
   with Weibull distribution it has been observed that the critical load $\sigma_c(N)$ for a bundle of size $N$ approaches to the 
   asymptotic values of 1/4 and $(5e)^{-1/5}$ \cite {Smith1982}. The numerical value of the finite size correction exponent $\nu$ has 
   been obtained very close to its exact value of 3/2 \cite {Smith1982,Daniels1989,Phoenix1992}. However the value of the exponent 
   $\kappa$ has been found to be slightly smaller than its exact value of $1/6$ \cite {Smith1982}. In addition following new results 
   have been obtained in this work. At the critical point the average relaxation time $\langle T(\sigma_c(N),N) \rangle$ grows as 
   $N^{\eta(N)}$ and the exponent $\eta(N)$ also approaches to its asymptotic value of 1/3. More importantly away from the critical 
   point the average relaxation time $\langle T(\sigma,N) \rangle$ obeys the usual scaling form with respect to $N$ and the deviation 
   from the critical point $|\Delta \sigma|$. Our most crucial result is we have not found any $\ln (N)$ dependence of the average
   relaxation time $\langle T(\sigma,N) \rangle$ in the precritical state.

      The research work in this paper is a part of the activity of the INDNOR (No: 217413/E20) project which is being acknowledged. 
   We thank Srutarshi Pradhan, B. K. Chakrabarti and Alex Hansen for important discussions.

   E-mail: manna@bose.res.in

\end {document}